\documentstyle[aps,prl]{revtex}
\begin{document}
\title{Deformation-induced thermomagnetic effects in a twisted
weak-link-bearing superconductor}
\author{Sergei A. Sergeenkov}
\address{Bogoliubov Laboratory of Theoretical Physics, Joint Institute for
Nuclear Research,\\ 141980 Dubna, Russia }
\address{\em (\today)}
\maketitle
\begin{abstract} Based upon the recently
introduced thermophase and piezophase mesoscopic quantum effects in
Josephson junctions, several novel phenomena in a twisted
superconductor (containing a small annular $SIS$-type contact) under
influence of thermal gradient and applied magnetic field are
predicted. Namely, we consider a torsional analog of Josephson
piezomagnetism (and related magnetomechanical effect) as well as a
possible generation of a heat flux induced magnetic moment in a
weakly-coupled superconductor under a torsional deformation (analog
of Zavaritskii effect) along with the concomitant phenomena of
piezothermopower and piezothermal conductivity. The conditions under
which the predicted effects can be experimentally measured in
conventional superconductors and nanostructured materials with
implanted Josephson contacts are discussed.
\end{abstract}

In 1972 Zavaritskii~\cite{1} observed for the first time a very
interesting phenomenon (the so-called deformation-induced
thermomagnetic effect): appearance of a heat flux $Q$ induced
magnetic field $\Delta H=H_q(\alpha )Q$ in rod-like tin samples (both
in the normal and superconducting state) under a torsional
deformation ${\cal M}$ (related to a torsional angle $\alpha ({\cal
M})={\cal M}/C_0$, where $C_0$ is a respective elastic modulus of the
material). Quite a tangible value of $\Delta H$ was registered under
the maximum load of ${\cal M}=0.2N/m$ (which corresponds to $\alpha
=0.01rad/cm$). This phenomenon was attributed to generation of
circular (non-potential) currents in a deformed sample (which in turn
lead to observable magnetic moments, see Ref.2 for discussion) and
further investigated by Lebedev~\cite{3} on the basis of the kinetic
theory.

At the same time, in response to fast growing interest to important
applications of Josephson and proximity effects in novel mesoscopic
quantum devices (such as, e.g., quantum computers), a substantial
progress has been made recently in measuring of (and manipulating
with) extremely small magnetic fields, thermal gradients and
mechanical deformations~\cite{4,5}.

Based upon the recently introduced thermophase~\cite{6,7} and
piezophase~\cite{8,9} effects (suggesting, respectively, a direct
influence of a thermal gradient and an applied stress on phase
difference through a Josephson junction), in this Letter we discuss
an analog of the above-mentioned Zavaritskii effect in a twisted
superconductor containing a single $SIS$-type contact and its
possible realization in conventional superconductors. Besides, we
also consider the concomitant phenomena of Josephson piezomagnetism
and magnetomechanical effect as well as the change of transport
properties of $SIS$-type junction under torsional deformation
(piezothermopower and piezothermal conductivity).

{\bf The model.} To follow the original paper of Zavaritskii~\cite{1}
as close as possible, let us consider a tin rod (of length $L$ and
radius $R$) with an anular $Sn-SnO-Sn$ contact~\cite{10} incorporated
into the middle of the rod (due to a high pliability of tin, it
should be quite easy to achieve), with a thin insulating $SnO$ layer
(of thickness $l$). Assuming a usual cylindrical geometry (with
$z$-axis taken along the rod's length and $A=\pi R^2$ being the
junction area), we can present the total Josephson energy on the
contact as follows (for the sake of simplicity, in this paper we
shall concentrate on zero-temperature effects only and will ignore
the role of Coulomb interaction effects assuming that the grain's
charging energy $E_c\ll E_J$, where $E_c=e^2/2C_J$, with $C_J$ being
the capacitance of the junction):
\begin{equation}
E_J=\int_0^\tau \frac{dt}{\tau}\int_{-R}^R\frac{dx}{A}
\int_{-\sqrt{R^2-x^2}}^{\sqrt{R^2-x^2}}dy \int_0^L\frac{dz}{L}[{\cal
H}(\vec x, t)]
\end{equation}
where the local Josephson energy is given by
\begin{equation}
{\cal H}(\vec x, t)=J\left [1-\cos \phi (\vec x, t)\right ]
\end{equation}
with the resulting phase difference
\begin{equation}
 \phi (\vec x, t)=\phi _0+\frac{2\pi dBx}{\Phi _0}
+\alpha ({\cal M})z+\frac{2\pi S_0\nabla T \vec x}{\Phi _0}t
\end{equation}
accounting for the change of the initial phase difference $\phi _0$
under the influence of an applied magnetic field $\vec B=(0,B,0)$,
thermal gradient $\nabla T=(\nabla _xT,0,\nabla _zT)$ and applied
torsional deformation ${\cal M}$ (through the corresponding torsional
angle $\alpha ({\cal M})$) taken along $z$-axis. Here $\Phi_ 0=h/2e$
is the quantum of flux with $h$ Planck's constant and $e$ the
electronic charge, $d=2\lambda _L+l$ is the junction size with
$\lambda _L$ being the London penetration length, $\tau$ is a
characteristic Josephson time~\cite{9}, $J$ is the Josephson coupling
energy, and $S_0$ is the field-free thermoelectric power (Seebeck
coefficient) on the junction.

The origin of the third term in Eq.(3) is quite obvious. Indeed,
under the influence of an homogeneous torsional deformation ${\cal
M}$, the superconducting phase difference will change with $z$ as
follows: $d\phi /dz=(d\phi /d\theta )(d\theta /dz)=N\alpha ({\cal
M})$ where~\cite{11} $\alpha ({\cal M})\equiv d\theta /dz={\cal
M}/C_0$ is the corresponding torsional angle variable and $N\equiv
d\phi /d\theta $ is a geometrical factor (in most cases~\cite{8}
$N\simeq 1$). As a result, the superconducting phase difference will
acquire an additional contribution $\delta \phi (z)=\alpha ({\cal
M})z$. (Notice that practically the same result can be obtained by
using the arguments from Ref.9 and invoking an analogy with a
conventional linear torsional piezoelectric effect which
predicts~\cite{12} $P({\cal M})=a{\cal M}$ for an induced electric
polarization.)

To neglect the influence of the self-field effects and ensure a
uniformity of the applied deformation ${\cal M}$ (and the related
torsional angle $\theta (L)\equiv \alpha ({\cal M})L$), we have to
assume that $\lambda _J>R$ and $L\gg R$ where $\lambda _J=\sqrt{\Phi
_0/2\pi d\mu _0j_c}$ is the Josephson penetration depth with $j_c$
being Josephson critical current density. As we shall see below,
these conditions can be reasonably well met experimentally.

{\bf Torsional piezomagnetic effect.} Before turning to the main
subject of this paper, let us briefly discuss the two preliminary
issues: (i) deformation induced behavior of the Josephson current and
(ii) torsional analog of Josephson piezomagnetism (which takes place
in a twisted $SIS$-type contact and manifests itself through
appearance of deformation-induced susceptibility) in the absence of
thermal gradient through the junction ($\nabla T=0$). Recalling the
definition of the Josephson current density $j_s(\vec x)=j_c\sin \phi
(\vec x, 0)$, in this particular case Eq.(3) brings about
\begin{equation}
I_s(B,\theta )=2I_c\left[\frac{J_1(B/B_0)}{B/B_0}\right
]\left(\frac{\sin \theta}{\theta}\right )
\end{equation}
for the maximum (with $\phi _0=\pi /2$) Josephson current in a
twisted cylindrical contact (under a torsional deformation ${\cal M}$
producing angle $\theta ={\cal M}L/C_0$) with $I_c=j_cA=2eJ/\hbar$.
Here $B_0=\Phi _0/2\pi dR$ is a characteristic Josephson field of
anular contact, and $J_1(x)$ is the Bessel function. Notice that as a
function of torsional angle $\theta$, the induced current
$I_s(B,\theta )$ follows a quasi-periodic Fraunhofer-like pattern and
reduces to the well-known~\cite{13} result for the magnetic field
dependence of an anular Josephson contact upon removal of mechanical
load (in deformation-free case when $\theta \to 0$).

Moving on to the second issue, we find that in addition to the
above-discussed angle-dependent Josephson current, in a twisted
contact will appear an induced magnetic moment (torsional
piezomagnetic effect)
\begin{equation}
M_s(B,\theta ) \equiv -\frac{1}{V}\left [\frac{\partial E_J}{\partial
B}\right ]_{\nabla T=0}
\end{equation}
where $V=AL$ is a sample's volume.

To capture the very essence of this effect, in what follows we assume
for simplicity that an {\it unloaded sample} does not possess any
spontaneous magnetization at zero magnetic field (that is
$M_s(0,0)=0$) and that its Meissner response to a small applied field
$B$ is purely diamagnetic (that is $M_s(B,0)\simeq -B$). According to
Eqs.(1)-(5), this condition implies $\phi _0=2\pi m$ for the initial
phase difference with $m=0,\pm 1, \pm 2,..$. As a result, we obtain
for the change of magnetization under torsional deformation
\begin{equation}
M_s(B,\theta )=-M_0f_1(B/B_0)g_0( \theta )
\end{equation}
Here $M_0=2J/VB_0$, $f_1(x)=-\frac{d}{dx}[f_0(x)]$ with
$f_0(x)=J_1(x)/x$, and $g_0(x)=\sin x/x$.

For low-field (Meissner) region, we can linearize the above equation
and define the deformation induced angle-dependent susceptibility
$\chi (\theta )$. Indeed, for $B\ll B_0$, Eq.(6) gives $M_s(B,\theta
)\simeq \chi (\theta )B$ where $\chi (\theta )=-\chi _0g_0(\theta )$
with $\chi _0=J/4VB_0^2$. As it follows from the above equations, the
superconducting (Meissner) phase of piezomagnetization $M_s(B,\theta
)$ (and the corresponding susceptibility $\chi (\theta )$) gradually
dwindles with increasing the angle $\theta $, shifting towards
paramagnetic phase (and reaching it eventually at $\theta \simeq
\pi$).

{\bf Magnetomechanical effect.} Let us consider the converse (to
piezomagnetism) magnetomechanical effect, that is field induced
change of torsional angle $\theta _s(B,{\cal M})$ (and corresponding
compliance $C_s^{-1}(B)$, see below). In view of Eqs.(1)-(3), we
obtain
\begin{equation}
\theta _s(B,{\cal M})\equiv \left [\frac{\partial E_J}{\partial {\cal
M}}\right ]_{\nabla T=0}=\theta _0f_0(B/B_0)g_1({\cal M}/{\cal M}_0)
\end{equation}
where $\theta _0=J/{\cal M}_0$ with ${\cal M}_0=C_0/L$ and
$g_1(x)=-\frac{d}{dx}[g_0(x)]$.

Notice that in the absence of applied magnetic field (when $B=0$) the
above equation establishes the so-called "torque-angle" relationship
(torsional analog of "stress-strain" law~\cite{8}) for a twisted
weak-link-bearing superconductor $\theta _s(0,{\cal M})=\theta
_0g_1({\cal M}/{\cal M}_0)$. For small enough torsional deformations
(when ${\cal M} \ll {\cal M}_0$ which is usually the case in
realistic experiments), the above model relationship reduces to a
more familiar Hooke's law, $\theta _s(0,{\cal M})={\cal M}L/C_s(0)$
with $C_s(0)=(3C_0/JL)C_0$ being the appropriate (zero-field) elastic
modulus (inverse torsional compliance) whose magnetic field
dependence is governed by the following equation
\begin{equation}
\frac{1}{C_s(B)} \equiv \frac{1}{L}\left [\frac{\partial \theta
_s(B,{\cal M})}{\partial {\cal M}}\right ]_{{\cal
M}=0}=\frac{1}{C_s(0)}\left[\frac{2J_1(B/B_0)}{B/B_0}\right ]
\end{equation}
It would be rather interesting to try to observe the above-discussed
piezomagnetic and magnetomechanical effects (including torsional
analog of paramagnetic Meissner effect~\cite{9}) in a twisted
weak-link-containing superconductor.

{\bf Deformation induced thermomagnetic effects.} Let us turn now to
the main subject of this paper and consider the influence of a
thermal gradient $\nabla T$ on the above discussed piezomagnetic and
magnetomechanical effects. Hereafter we restrict our consideration to
the case of small values of the applied thermal gradient $\nabla T$,
leading to linear thermoelectric effects (for discussion of possible
nonlinear Seebeck effects in Josephson junctions and granular
superconductors see Ref.7). For thermal gradient applied normally and
parallel to the torsional deformation (see Eq.(3)), we obtain two
contributions (transverse and longitudinal) for the deformation
induced thermomagnetization, emerging in the vicinity of the
Josephson contact
\begin{equation}
\Delta M(\theta ,B,\nabla T) \equiv -\frac{1}{V}\left [\frac{\partial
E_J}{\partial B}\right ]_{\nabla T\neq 0} =M^q_{\bot}(\theta
,B)\nabla _xT + M^q_{\|}(\theta ,B)\nabla _zT
\end{equation}
where
\begin{equation}
M^q_{\bot}(\theta ,B)=M^q_{0\bot}f_2(B/B_0)g_0(\theta )
\end{equation}
and
\begin{equation}
M^q_{\|}(\theta ,B)=M^q_{0\|}f_1(B/B_0)g_1(\theta )
\end{equation}
Here $M^q_{0\|}=(L/R)M^q_{0\bot}$ with $M^q_{0\bot}=2eJS_0\tau R/B_0V
\hbar$, and $f_2(x)=\frac{d}{dx}[f_1(x)]$.

Notice that within the geometry adopted in the present paper, the
true analog of Zavaritskii effect is given by a zero field limit of
the transverse (paramagnetic) contribution, $M^q_{\bot}(\theta
,0)=(1/8)M^q_{0\bot}(\sin \theta /\theta )$. Its evolution with
torsional angle follows the corresponding behavior of the induced
susceptibility $\chi (\theta )$ (Cf. Eq.(6)) and, thus, it also
undergoes a "diamagnetic-paramagnetic" transition upon reaching a
critical angle $\theta _c \simeq \pi$. On the other hand, the
field-dependent longitudinal component $M^q_{\|}(\theta ,B)$ changes
with $\theta$ linearly for $\theta \ll 1$. It describes the
appearance of deformation induced thermomagnetic component normal to
the applied magnetic field $\vec B=(0,B,0)$. It disappears when $B
\to 0$ and for non-zero fields it closely follows the behavior of the
(initially diamagnetic) torsional piezomagnetization $M_s(B,\theta )$
considered in the previous section (Cf. Eq.(6)).

Likewise, for thermal gradient applied normally and parallel to the
torsional deformation (see Eq.(3)), we obtain two contributions for
the heat flux induced magnetomechanical effect
\begin{equation}
\Delta \theta ({\cal M} ,B,\nabla T) \equiv \left [\frac{\partial
E_J}{\partial {\cal M}}\right ]_{\nabla T\neq 0}
 =\theta ^q_{\bot}({\cal M}
,B)\nabla _xT + \theta ^q_{\|}({\cal M} ,B)\nabla _zT
\end{equation}
where
\begin{equation}
\theta ^q_{\bot}({\cal M} ,B)=\theta ^q_{0\bot}f_1(B/B_0)g_1({\cal
M}/{\cal M}_0)
\end{equation}
and
\begin{equation}
\theta ^q_{\|}({\cal M} ,B)=\theta ^q_{0\|}f_0(B/B_0)g_2({\cal
M}/{\cal M}_0)
\end{equation}
Here $\theta ^q_{0\|}=(L/R)\theta ^q_{0\bot}$ with $\theta
^q_{0\bot}=2eS_0JR\tau /\hbar {\cal M}_0$ and
$g_2(x)=-\frac{d}{dx}[g_1(x)]$.

Once again, the true (deformation-free) heat flux induced
magnetomechanical effect is given by the zero-deformation limit of
the longitudinal component $\theta ^q_{\|}(0,B)$ while a small angle
expansion of the transverse component describes a thermal analog of
Hooke's law $\theta ^q_{\bot}({\cal M} ,B)\simeq {\cal M}/C
^q_{\bot}(B)$ with $C^q_{\bot}(B)$ being the appropriate compliance
coefficient.

{\bf Piezothermopower.} Let us briefly discuss now one more
interesting phenomenon of {\it piezothermopower} which can occur in a
twisted superconducting rod with a $SIS$-type junction. According to
Eqs.(1)-(3), the transverse and longitudinal contributions to
(magnetic field-dependent) change of junction's Seebeck coefficients
under torsional deformation read:
\begin{equation}
\Delta S_{\bot}(\theta ,B) \equiv -\frac{1}{eR}\left [\frac{\partial
E_J}{\partial \nabla _x T}\right ]=S_Jf_1(B/B_0)g_0(\theta )
\end{equation}
and
\begin{equation}
\Delta S_{\|}(\theta ,B) \equiv -\frac{1}{eL}\left [\frac{\partial
E_J}{\partial \nabla _z T}\right ]=-S_Jf_0(B/B_0)g_1(\theta )
\end{equation}
with $S_J=(2\tau J/\hbar )S_0$.

Like in the previous paragraph, the deformation-free transversal
component $\Delta S_{\bot}(0,B)$ describes the evolution of
conventional thermopower in unloaded ($\theta =0$) sample with
applied magnetic field, while the true piezothermopower is given by a
zero-field limit of the longitudinal component $\Delta S_{\|}(\theta
,0)$. Besides, the above analysis allows us to introduce a
deformation induced AC Josephson effect (as a generalization of a
more familiar thermal AC effect~\cite{14}) in a twisted $SIS$-contact
bearing superconductor. Indeed, as soon as the thermal current
exceeds the critical current, which will happen when $\Delta T$
becomes larger than $\Delta T_c(\theta ,B)=I_cR_n/\Delta
S_{\|}(\theta ,B)$ (where $R_n$ is the normal state resistance), a
new type of Josephson generation with frequency $\omega (\theta
,B)=2e\Delta S_{\|}(\theta ,B)\Delta T/\hbar$ should occur in the
junction area (see below for more discussion).

{\bf Piezothermal conductivity.} Finally, let us briefly consider the
influence of mechanical deformation on magnetothermal conductivity of
the twisted weak-link-bearing superconducting rod. To this end, we
recall that the local heat flux density $\vec q(\vec x,t)$ is related
to the local Josephson energy density ${\cal H}(\vec x,t)$ via the
energy conservation law
\begin{equation}
\nabla \vec q(\vec x,t)+{\dot{\cal H}}(\vec x,t)=0
\end{equation}
where ${\dot{\cal H}}=\partial {\cal H}/\partial t$.

The above equation allows us to introduce an effective thermal flux
\begin{equation}
\vec Q(t)\equiv \frac{1}{V}\int d^3x\vec q(\vec x,t)=\frac{1}{V}\int
d^3x {\dot{\cal H}}(\vec x,t)\vec x
\end{equation}
which in turn is related to the thermal conductivity tensor $\kappa
_{\alpha \beta}$ as follows ($\{ \alpha ,\beta \}=x,y,z$)
\begin{equation}
<Q_{\alpha}>\equiv \frac{1}{\tau}\int_0^\tau
dtQ_{\alpha}(t)=-\kappa _{\alpha \beta}\nabla _{\beta}T
\end{equation}
Straightforward calculations based on Eqs.(1)-(3) yield the following
explicit expressions for non-zero components of the thermal
conductivity tensor
\begin{eqnarray}
\kappa _{xx}(\theta ,B)&=&\kappa _0\left(\frac{R}{L}\right
)J'_1(B/B_0)g_{xx}(\theta )\\  \nonumber \kappa _{xz}(\theta
,B)&=&\kappa _{zx}(\theta ,B)=\kappa _0J_1(B/B_0)g_{xz}(\theta ) \\
\nonumber \kappa _{zz}(\theta ,B)&=&\kappa _0\left(\frac{L}{R}\right
)J_0(B/B_0)g_{zz}(\theta ) \nonumber
\end{eqnarray}
where $\kappa _0=8eS_0JRL/hV$, $J'_1(x)=J_0(x)-J_1(x)/x$ with
$J_n(x)$ being the corresponding Bessel functions, $g_{xx}(x)=(1-\cos
x)/x$, $g_{xz}(x)=\frac{d}{dx}[g_{xx}(x)]$, and
$g_{zz}(x)=-\frac{d^2}{dx^2}[g_{xx}(x)]$.

For small torsional deformations (with $\theta \ll 1$),
$g_{xx}(\theta )\simeq \theta /2$, $g_{xz}(\theta )\simeq (1-\theta
^2/4)/2$ and $g_{zz}(\theta )\simeq \theta /4$. So, $\kappa
_{xz}(0,B)$ describes a conventional (deformation-free)
magnetothermal conductivity in unloaded junction~\cite{15}, while
true (magnetic field-free) piezothermal conductivity is given by
$\kappa _{xx}(\theta ,0)$ and $\kappa _{zz}(\theta ,0)$ components
(both increasing linearly with $\theta$ for small angles). In a
sense, introduced here piezothermal conductivity is a converse effect
with respect to the original Zavaritskii effect~\cite{1,2} and it
seems very interesting to try to realize it experimentally using tin
rods (with and without weak links).

{\bf Discussion.} To estimate the magnitudes of the predicted
effects, we make use of the fact that for thermoelectric processes
the characteristic time $\tau$ is related to a thermal AC frequency
$\omega _T=2eS_0\Delta T/\hbar$ (assuming $\nabla _xT \simeq \Delta
T/R$ and $\nabla _zT \simeq \Delta T/L$) as follows: $\tau =2\pi
/\omega _T$. As a result, the deformation induced thermomagnetic
field (Josephson analog of Zavaritskii effect) reads:
\begin{equation}
\Delta H_{\bot}\equiv 4\pi M^q_{\bot}(\theta ,0)\nabla _xT\simeq
H_J\left(\frac{R}{L}\right )\left(\frac{\sin \theta}{\theta}\right )
\end{equation}
where $\mu _0H_J=\Phi _0/2\pi \lambda _J^2$ is a critical Josephson
field.

Furthermore, combining the experimental parameters for $Sn-SnO-Sn$
anular Josephson junction considered by Matisoo~\cite{10} (namely,
$R_n=10^{-9}\Omega$, $j_c(0)=10^4A/m^2$, $\lambda _L(0)=40nm$, and
$\lambda _J(0)=5mm$) with the typical parameters from Zavaritskii
experiments on tin rods~\cite{1,2} (namely, $2R=5mm$, $L=8cm$, and
$\theta _{max}=0.1 rad$) we obtain $\Delta B_{\bot}=\mu _0 \Delta
H_{\bot} \simeq 10^{-12}T$ which is equivalent to quite a tangible
value of the deformation induced thermomagnetic flux through the
junction $\Delta \Phi _{\bot}=\pi R^2\Delta B_{\bot} \simeq
10^{-2}\Phi _0$. The same set of parameters yields $B_0=\Phi _0/2\pi
dR\simeq 10^{-6}T$ and $B_J=\Phi _0/2\pi \lambda _J^2 \simeq
10^{-10}T$ for estimates of characteristic and critical Josephson
fields, respectively. Finally, using some typical~\cite{14}
experimental data describing physics of conventional Josephson
contacts (with $S_0\simeq 10^{-8}V/K$ and $\Delta T\simeq 10^{-4}K$),
we can estimate the orders of magnitude of the piezothermal
conductivity and deformation induced AC Josephson generation (related
to piezothermopower). The result is as follows, $\kappa _{zz}(\theta
,0)\simeq (2eS_0JL^2/hV)\theta \simeq (2/\pi )S_0j_c\theta \simeq
10^{-2}W/mK$ and $\omega (\theta ,0)=2e\Delta S_{\|}(\theta ,0)\Delta
T/\hbar \simeq (\omega _J\tau /3)\theta \omega _T \simeq 10^{11}Hz$,
where $\omega _J=2J/\hbar \simeq 10^{12}Hz$ and $\omega
_T=2eS_0\Delta T/\hbar \simeq 1kHz$ are characteristic (Josephson and
thermal, respectively) frequencies. The above estimates suggest quite
an optimistic possibility to observe the predicted here effects
experimentally through a comparative study of conventional
superconductors with and without $SIS$-type junctions.

At the same time, to check the validity of some other interesting
predictions (like Fraunhofer patterns given by a set of $g_n(\theta
)$ functions, see the text), much larger torsional deformations
(reaching critical angles of the order of $\theta _c\simeq \pi$) are
needed. Hopefully, this will become possible in the very nearest
future, with further advancement of experimental techniques and new
technologies for manufacturing of nanostructured superconducting
materials with implanted atomic scale Josephson junctions and other
weak links~\cite{4}.

In conclusion, a few novel mesoscopic quantum phenomena which are
expected to occur in a weak-link-bearing superconductor under
influence of torsional deformation, thermal gradient and applied
magnetic field were presented. The observability of the predicted
effects, using conventional superconductors as well as novel
materials with well-controlled mesoscopic Josephson junctions was
discussed.

The idea of this work was conceived and partially realized during my
stay at the Universidade Federal de S\~ao Carlos (Brazil) where it
was funded by the Brazilian Agency FAPESP (Projeto 2000/04187-8). I
thank Wilson Ortiz and Fernando Araujo-Moreira for hospitality and
stimulating discussions on the subject.

\end{document}